\documentclass[mathleft
]{an}
\usepackage{graphicx}
\usepackage{times}
\usepackage{cite}
\usepackage{graphicx}
\usepackage{amssymb}
\usepackage[dvipsnames]{color}

\begin{document}

\Pagespan{625}{631}
\Yearpublication{2008}%
\Yearsubmission{2007}%
\Month{11}%
\Volume{329}%
\Issue{6}%
\DOI{10.1002/ansa.200711001}%

\newcommand{\3}{\ss}
\newcommand{\n}{\noindent}
\newcommand{\eps}{\varepsilon}
\newcommand{\be}{\begin{equation}}
\newcommand{\ee}{\end{equation}}
\def\ba{\begin{eqnarray}}
\def\ea{\end{eqnarray}}
\def\de{\partial}
\def\msun{M_\odot}
\def\div{\nabla\cdot}
\def\grad{\nabla}
\def\rot{\nabla\times}
\def\ltsima{$\; \buildrel < \over \sim \;$}
\def\simlt{\lower.5ex\hbox{\ltsima}}
\def\gtsima{$\; \buildrel > \over \sim \;$}
\def\simgt{\lower.5ex\hbox{\gtsima}}

\title{Low-temperature primordial gas in merging halos}

\author{E.~O.~Vasiliev\inst{1,2}\fnmsep\thanks{Corresponding author:
  \email{eugstar@mail.ru}}
          \and
Yu.~A.~Shchekinov\inst{3,4}
}

\titlerunning{Low-temperature primordial gas}
\authorrunning{E.~O.~Vasiliev \& Yu.~A.~Shchekinov}
   
   \institute{Tartu Observatory, 61602 T\~oravere, Estonia 
\and
Institute of Physics, Southern Federal University,
Stachki St. 194, Rostov-on-Don, 344090 Russia
\and
Department of Physics, Southern Federal University,
Sorge St. 5, Rostov-on-Don, 344090 Russia
\and 
Special Astrophysical Observatory, Nizhny Arkhyz, 
Karachaevo-Cherkessia, 369167 Russia  
}

\received{11 May 2007}
\accepted{25 Apr 2008}
\publonline{2008 Jun 25}

\keywords{cosmology: early universe -- galaxies: formation --
ISM: molecules -- stars: formation -- shock waves}

\abstract
{Thermal regime of the baryons behind shock waves arising
in the process of virialization of dark matter halos is governed
at cetrain conditions by radiation of HD lines. A small fraction of
the shocked gas can cool down to the temperature
of the cosmic microwave background (CMB). We 
estimate an upper limit for this fraction: at $z=10$ it increases sharply
from about $q_{_T}\sim 10^{-3}$ for 
dark halos of $M=5\times 10^7\msun$ to
$\sim 0.1$ for halos with $M=10^8\msun$. Further
increase of the halo mass does not lead however to a significant
growth of $q_T$ -- the asymptotic value for $M\gg 10^8\msun$ is of 0.3:
We estimate star formation
rate associated with such shock waves, and show that they can provide a
small but not negligible fraction of the star formation.
We argue that extremely metal-poor low-mass stars in the
Milky Way may have been formed from primordial gas behind such shocks.
}

\maketitle

\section{Introduction}

Formation of the first stars is determined by energy losses in ro-vibrational
lines of molecular hydrogen H$_2$ and its deuterated analogue HD
(Lepp \& Shull 1983; Shchekinov 1986; Puy et al. 1993; Palla, Galli \& Silk 1995;
Galli \& Palla 1998, 2002; Stancil, Lepp \& Dalgarno 1998; Tegmark et al. 1997;
Puy \& Signore 1997, 1998; Uehara \& Inutsuka 2000; Flower 2002; Nakamura \&
Umemura 2002; Machida et al. 2005, Nagakura \& Omukai 2005).
In turn, the amount of H$_2$ and HD and their ability to cool gas greatly depend on
dynamical and thermal regime of the gas. In particular, shock waves
strongly enhance the rate of conversion of atomic hydrogen
to its molecular form
(Shchekinov \& Entel 1983; Suchkov, Shchekinov \& Edelman 1983; Shapiro \& Kang 
1987; Kang \& Shapiro 1992; Shchekinov 1991; Ferrara 1998; Yamada \& Nishi 1998;
Uehara \& Inutsuka 2000; Cen 2005; Machida et al. 2005; Shchekinov \& Va\-si\-liev
2005; Johnson \& Bromm 2005).
On the other hand, the first stars have formed from the gas processed 
by the shock waves unavoidable in the process of virialization of dark matter halos
(Shapiro 1993; Haiman, Thoul \& Loeb 1996; Tegmark et al. 1997; Abel, Bryan \& 
Norman 2000, 2002).
Therefore possible enhancement of H$_2$ and
HD in these conditions can have important consequences for characteristics
of the first stars (Oh \& Haiman 2002; Johnson \& Bromm 2005; Shchekinov \& Vasiliev 2005).

When dark matter halos merge, shock waves form and compress the baryons
(see discussion in (Barkana \& Loeb 2001; Ciardi \& Ferrara 2004)).
At sufficiently large velocities of colliding flows
($v>8$ km s$^{-1}$) fractional ionization in shocked gas increases above
the frozen cosmological value, and speeds up chemical
kinetics of H$_2$ molecules. Collisions with velocities
above $v>8.6[(1+z)/20]^{-1/6}$ km s$^{-1}$ may lead to a
rapid formation of HD and an efficient cooling down to the minimum
temperature $T=T_{\rm CMB}=2.7(1+z)$ 
(Vasiliev \& Shchekinov 2005; Johnson \& Bromm 2005; Shchekinov \& Vasiliev 2005).
It is worth noting that HD can also play major role 
in central regions of collapsing dark halos, where a high-density 
environment favours transition of D into molecular form, so that 
even in low-mass halos HD cooling can dominate (Ripamonti 2007).

Explicit answer of how
large is the fraction of cold gas in merging halos
can be found only in hydrodynamic
simulations. Very recently (Maio et al. 2007)
performed 3D SPH simulation of structure formation with accounting of HD chemistry. 
They found that incorporation of effects from HD cooling increases 
clumpiness of gas, i.e. clouds are den\-ser
and more compact with respect to the case when only H$_2$ cooling is
considered. However,
full 3D simulations are time consuming,
and always are made within a particular realization of a
random hydrodynamic field. They therefore represent only very limited 
regions in the
space of possible random hydrodynamic fields of a given spectrum.
This circumstance restricts the final thermodynamic
state of bary\-ons within the domain corresponding to the chosen interelations
between the amplitudes of different wave modes. As a result, estimates of the
fraction of the cold baryons able to form stars are biased by such limitations.
At such circumstances 
1D hydrodynamical simulations of chemistry and thermodynamics of
shocked primordial gas can play an auxiliary role for understanding of what in principle can
be expected in the conditions preceeding formation of the first stars.
In this paper we use 1D planar computations to model chemical and thermal regime of
gas behind the shock waves after a head-on collision of two clouds of equal sizes. This approach is justified by the fact that in supersonic cloud collisions the
rarefaction motion transverse to the symmetry axis takes longer  than the
collision time $t_c=3R/2v_c$ (Gilden 1984). On the other 
hand, the estimates of the fraction of baryons cooled down by HD molecules we obtaine within this approximation are obviously upper limits.

At the stages when the halos are close to merge their velocities are of the order
of the virial value corresponding to the total mass of the halos. We
therefore connect the collisional velocities of the clouds to
the mass of the halo $M$ formed in this collision
\begin{equation}
\label{vel}
v_c=(3^4\pi^3\Omega_m\rho_c)^{1/6}G^{1/2}M^{1/3}(1+z)^{1/2},
\end{equation}
$\Omega_m$ is the matter closure parameter, $\rho_c=3H_0^2/8\pi G$, the
critical density.

In Section 2 we describe dominant thermal and chemical processes in shocked gas;
in Section 3 we discuss consequences of HD cooling for formation
of the first stars; summary is given in Section 4.

Throughout the paper we assume a $\Lambda$CDM cosmology with the parameters
$(\Omega_0, \Omega_{\Lambda}, \Omega_m, \Omega_b, h ) = 
(1.0,\ 0.7,\ 0.3,\\ 0.045,\ 0.7 )$
and deuterium abundance $2.6\times 10^{-5}$,
consistent with the most recent measurements (Spergel et al. 2007).

\section{Chemistry and thermal regime behind the shock}

In the center of mass of the colliding baryon components
a discontinuity forms at the symmetry plane, and two shock waves
begin to move outwards. We assume that during the merging
collisionless dark matter components
occupy considerably bigger volume and neglect gravitational
forces on baryons. Therefore we describe propagation of the shock
by single-fluid hydrodynamic equations. Energy equation includes
radiative losses typical in primordial plasma:
Compton cooling, recombination and
bremsstrahlung radiation, collisional excitation of HI (Cen 1992)
H$_2$ (Galli \& Palla 1998)
and HD (Flower 2000; Lipovka, N\'un\~ez-Lo\-p\'ez \& Avi\-la-Ree\-se 2005). 
More accurately The H$_2$ cooling function have been recalculated by 
Le Bourlot, Pineau des Forets \& Flo\-wer (1999), which 
which differs considerably (by factor of 2) from the Le Bourlot, 
Pineau des Forets \& Flo\-wer function only at $T>4000$~K (Glover 2005).

Chemical and ionization
composition include a standard set of species: H, H$^+$, $e$, H$^-$, He,
He$^+$, He$^{++}$, H$_2$, H$_2^+$, D, D$^+$, D$^-$, HD, HD$^+$; 
The corresponding rates are taken from (Galli \& Palla 1998; Stancil et al. 1998);
the shock wave was computed on one collision time $t_c$. 
H$_2$ dissociation rate by atomic hydrogen is taken from 
Mac Low \& Shull (1986) -- although this reaction is important only 
at $T>6000$ K, it restricts the production of H$_2$ molecules by 
factor of 1.5.

For our simulations we use a 1D planar  Lagragian
sche\-me similar to that described by Thoul \& Weinberg (1995). As a 
standard resolution we have used 700 zones over the computational region, 
and found a resonable convergence: several test computations with 1500 
zones showed significant deviations only in the very central zones 
of 1\% in mass, while in the rest the deviation is normally less than 
10 \%. 
We assume a ``top-hat'' initial baryonic distribution
in colliding halos with the density equal to the virialized value
$18\pi^2 \Omega_b \rho_0 (1+z)^3$, while temperature is taken close
to the cosmic microwave background (CMB)
temperature $T_b=1.1 T_{\rm CMB}$. This corresponds to a
simplified picture when merging halos are already compressed to their
virial radii, but not yet virialized, i.e. systematic large scale motions did not converged into thermal energy. 
This assumption allows to better
understand the role which shock compression of the
baryons plays in their ionization
and chemical state in the course of virialization.
The fractional ionization $x$, and the abundances of H$_2$ and HD
molecules before the shock are taken equal to their background values
$x=10^{-4}$, $f({\rm H}_2)=10^{-6}$ and $f({\rm HD})=10^{-10}$.

Fig. 1 shows typical distributions of the temperature, the abundances
of H$_2$ and HD and their relative contributions to the total cooling behind the
shock front at $t=(0.2,~0.6,~1)t_c$,
for the halos merged at $z=20$ with the velocity $v_c=20$ km s$^{-1}$, corresponding to
the total mass $M=1.4\times 10^7M_\odot$. Three qualitatively different cooling
regimes can be distinguished in the temperature profiles: in the high temperature
range ($1500<T<7000$ K) excitation of ro-vibrational levels of H$_2$ dominates,
while in the intermediate range ($200<T<1500$ K) only rotational lines contribute
to the cooling; in the lowest temperature range ($T<200$ K) cooling
from H$_2$ molecules exhausts and only HD rotations support cooling -- it is
seen in the lower panel from comparison of the relative
contributions of H$_2$ and HD cooling. At later times, 
$t\sim t_c$ a nonmonotonous behaviour of temperature is due to the fact 
that at these stages gas enters the shock front with lower velocities, 
so that the post-shock temperature becomes below the limit when 
collisional dissociation of H$_2$ molecules is important. As a result, 
all molecular hydrogen formed behind the front survives and stimulates 
rapid cooling. 

\begin{figure}
  \resizebox{\hsize}{!}{\includegraphics{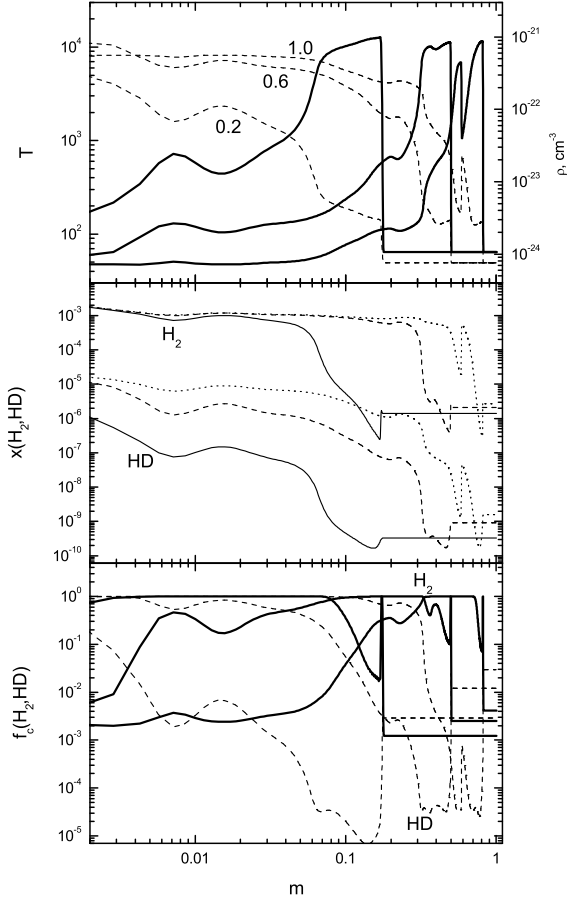}}
      \caption{
      Upper panel: profiles of temperature (solid) and density (dash);
      middle panel: relative concentration of H$_2$ and HD molecules;
      lower panel: their relative contribution to the total cooling
      (H$_2$ -- solid and HD -- dashed), for
      baryons in two colliding halos with the total mass $M=1.4\times 10^7M_\odot$
      at $0.2t_c,~0.6t_c,~t_c$; halos merged at $z=20$.
                }
  \label{Fig1}
\end{figure}

In this particular case a small fraction ($q\simeq 0.1$ by mass) of
the shocked baryons close to the symmetry plane
cools to the minimum possible value $T\simeq T_{\rm CMB}=2.7(1+z)$ due to
cooling in rotational lines of HD. In general, the fraction of compressed baryons
cooled down to a given temperature
level depends on the relative velocities of the colliding
clouds $v_c$: the larger the collisional velocity $v_c$,
the stronger the gas compression after
cooling, and the higher the contribution from HD cooling.
Fig. 2 shows the fraction of baryons $q_T(M)$ contained in several temperature ranges
versus the halo mass:
$T<200$ K, $T<150$ K, $T<100$ K, and at $T\simeq T_{\rm CMB}$ after one crossing
time $t=t_c$.
In the temperature range $T<200$ K where cooling in H$_2$ lines is dominant,
$q_T(M)$ increases from 0.1 for the halo mass $10^7 M_\odot$ to the asymptotical
(at $M\gg 2\times 10^7 M_\odot$) value 0.6.
Compressed baryons can have temperature below
150 K only due to contribution from HD cooling. In the lower temperature ranges
($T<100$ K, and at $T=T_{\rm cmb}$) $q_T(M)$ is a very sharp function of
the halo mass: for instance, at redshift $z=20$ (Fig. 2a)
a two-fold increase of the mass from $10^7
M_\odot$ to $2\times 10^7M_\odot$ results in a two-order of magnitude
increase of $q_{T_{\rm CMB}}$ from $3\times 10^{-3}$ to 0.3. At higher masses the
dependence flattens and asymptotically in the limit $M\gg 2\times 10^7M_\odot$
approaches $0.4$. At lower redshifts gas density decreases and
the radiation cooling time becomes longer. As a result,
$q_T(M)$ shifts towards larger halo
masses, approximately as $(1+z)^{-3/2}$: Fig. 2b shows $q_T(M)$ for collisions
occured at $z=10$.

\begin{figure}
  \resizebox{\hsize}{!}{\includegraphics{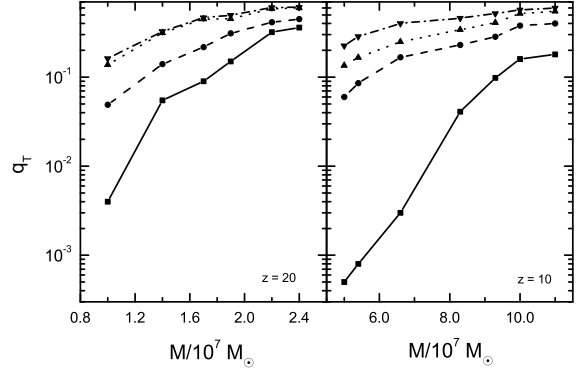}}
      \caption{
      Fraction of baryons cooled below temperature $T<200$ K,
      $T<150$ K, $T<100$ K, and $T=T_{\rm CMB}$ from top to bottom at
      $z = 20$ (left), and $z = 10$ (right).
                }
  \label{Fig2}
\end{figure}

\section{Star formation}

Baryons cooled below $T<200$ K are thought to be able to
fragment and initiate star formation. In gas layers compressed by shock waves
gravitational instability of the cold gas occurs naturally when the thickness
of the layer equals the Jeans length. In order to estimate the range of
masses expected to form through the gravitational instability we apply
(Gilden 1984)
criterion for the shock-compressed gas, which implies that:
{\it i)} the characteristic growth time is short\-er
than the collision time, and {\it ii)} the critical wavelength is shorter
than the initial size of the clouds. The corresponding critical mass
$M_{\rm cr}$ depends on the average temperature and density in the layer:
when the halos merge with low relative velocities
(corresponding to lower halo masses), only a
small fraction of the compressed baryon mass can cool down to sufficiently
low temperatures to form an unstable layer. Mergings with higher
relative velocities increase the fraction of gravitationally unstable baryons.
Fig.3 shows the redshift dependence of the halo masses with a given fraction of the
compressed baryons unstable in Gilden sense.
Each line is marked with symbols corresponding to a given
fraction of the baryon mass
$q_f$ unstable against fragmentation: for instance, when halos with masses
corresponding to the upper line $M=2.5\times 10^7[(1+z)/20]^{-2.2} M_\odot$ merge,
half of their mass becomes compressed in a cold layer
with temperature  $T\leq 100-150$ K unstable in Gilden sense. 
 Coincidently, the thermal evolution is governed by HD cooling,
when the temperature falls below 150~K.
At the latest stages the unstable layers are dominated by HD cooling, 
so that fragments formed in these conditions can reach the minimum possible
temperature $\simeq T_{\rm CMB}$. The corresponding Jeans mass in the unstable
layer is $M_J \leq 2.3\times 10^3 M_\odot v_{10}^{-1} [(1+z)/20]^{1/2}$,
which is considerably smaller than the baryonic mass of this layer;
here $v_{10} = v_c/10$ km s$^{-1}$. With accounting (\ref{vel})
the Jeans mass in the unstable layer is $M_J\leq
1.3\times 10^4M^{-1/3}M_\odot$. This means that when halos
with masses $M>10^4M_\odot$ merge more than one cold and dense
clouds in the unstable layer can form and give rise to formation
of stars.
\begin{figure}
  \resizebox{\hsize}{!}{\includegraphics{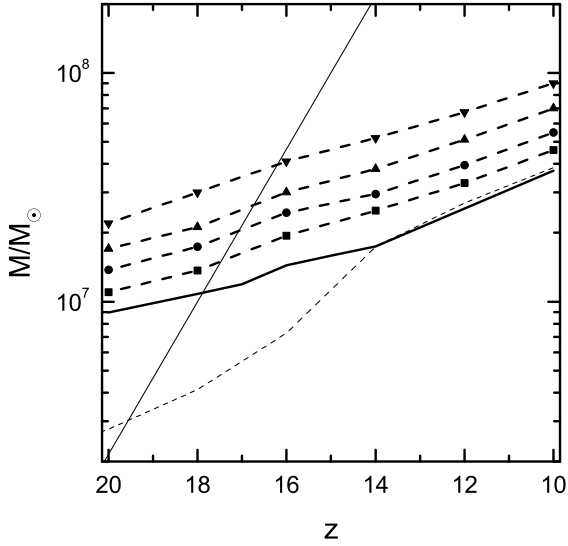}}
      \caption{
      Lines with simbols depict halo masses vs redshift, where
      the fraction of baryon mass unstable in Gilden sense is at given level:
      $1\%,\ 7\%,\ 25\%,\ 50\%$ from bottom to top; straight thin solid line
      corresponds to a 3$\sigma$ peak mass, dashed line shows the minimum
      mass obtained by (Tegmark et al. 1997), thick solid line shows the
      minimum mass from (Shchekinov \& Vasiliev 2006).
      }
   \label{Fig3}
\end{figure}

Whether such fragments are the protostellar condensations evolving further
in a single massive star, or they are the nestles, where in the process of
sequential (hierarchical, Hoyle 1953)
or of a one-step (Nakamura \& Umemura 1999)
fragmentation a cluster of less massive
stars is form\-ed, depends sensitively on details of gravitational contraction
and thermal evolution of the fragments
(see discussion in (Coppi, Bromm \& Larson 2001; Glover 2004).
However, both
the mass of a central protostellar core in the former case, and the minimum
mass of a protostellar condensation in the latter are determined by
the opaqueness of the contracting gas.
Vasiliev \& Shchekinov (2005) estimate this mass as
$M_J\sim 10^{-3}(1+z)^{3/2}M_\odot$, what gives for
$z=10-20$ a relatively low mass limit: $M_{\rm J}\sim (0.03-0.1)M_\odot$.
Therefore stars formed in such cold layers with a predominance of HD
cooling on later stages, are anticipated in general
to be less massive than those expected
when thermodynamics of primordial gas is determined by
H$_2$ cooling (Shchekinov \& Vasiliev 2005).

The fraction of baryons in the universe able to cool below 
 $T=150$ K and to form
presumably low-mass stars behind shock waves in mergings can be estimated as
\begin{eqnarray}
\label{massint}
f_c=\epsilon_\perp\left.\int\limits_{M_{\rm c}}^{M_{4\sigma}}
q_{_{T\leq 100}}(M,z)MF(M)\,dM\right/ \nonumber \\
\int\limits_{M_{\rm min}}^{M_{4\sigma}}MF(M)\,dM,
\end{eqnarray}
where $F(M)=dN/dM$ is the Press-Schechter
mass function, $M_{\rm c}$ is the halo mass at a given redshift 
presumably formed through merging of two $M_{\rm c}/2$ halos, 
where the fraction of cold ($T\leq 150$ K) baryons is equal 
to 1\%, $M_{\rm min}$, the minimum halo mass at a given redshift. 
We assume that the halos with the fraction of cold baryons
less than 1\% do not significntly contribute to $f_c$ or the star formation rate.
Here we introduced factor $\epsilon_\perp=0.05$ accounting
only approximately head-on collisions. Indeed,
our conclusions about the role of HD cooling
are based on the assumption of a head-on
collision of merging halos, and can be valid only in a restricted range of the
impact parameter when the shear motion is less important than the converging flow
and the corresponding diverging shock waves. For this condition to be
fulfilled the
characteristic time of the Kelvin-Helmholtz instability of the shear flow
$t_{_{\rm KH}}\sim R/v_{||}$
must be longer than the dynamical (crossing) time $t_{d}\sim 2R/v_\perp$:
$t_{_{\rm KH}}>t_{d}$. This gives
$v_{||}/v_\perp<1/2$, and as a result, only a fraction $\epsilon_\perp=
\Delta \Omega/4\pi\simeq 0.05$ of mergers where the flows are approximately head-on. 
In Introduction we stressed that the estimate of the cold baryon fraction 
$f_c$ within the approximatrion of head-on collisions is an upper limit.  Introducing the factor $\epsilon_\perp$ attempts to quantify more 
accurately the fraction $f_c$. However, the estimate (\ref{massint}) 
still has a meaning of an upper limit, particularly because {\it i)} 
in (\ref{massint}) we assumed explicitly that the spectrum $F(M)$ above 
$M_c$ is formed thro\-ugh merging of equal masses, and {\it ii)} 
we do not 
account here (negative) feedback from the stars have already formed 
during the merging. 

Fig. 4 depicts $f_c$ versus redshift. 
For $M_{\rm min}$ we have taken $10^3M_\odot$,
$10^4M_\odot$ and $10^5M_\odot$, however, only the last value seems meaningful,
because for lower halo masses the\-ir baryonic content is too small and
likely can be easily removed in tidal interactions. Independent on
$M_{\rm min}$ the total fraction of baryons in the universe
able to cool below $T=150$ K and form stars approaches
$f_c\sim 0.1\epsilon_\perp$ at $z=10$. This fraction may increase if the halos
are clustered as suggested by Barkana \& Loeb (2004) and  Cen (2005).
For comparison we show in Fig. 4 the
fraction of baryons in the universe cooled by
H$_2$ molecules and hydrogen atoms with 
the lower limit $M_{\rm c}$ in the integral (\ref{massint}) 
corresponding to the virial mass $M_{\rm c}=M_v$ at 
$T_{vir} = 400,~10^4$~K, for both cases
we assume that in virialized halos 8\% of baryons can
have temperature 200 K as calculated by (Abel et al. 1998),
and integrate over the halo mass spectrum from $M_{\rm min}=10^4~M_\odot$.

\begin{figure}
  \resizebox{\hsize}{!}{\includegraphics{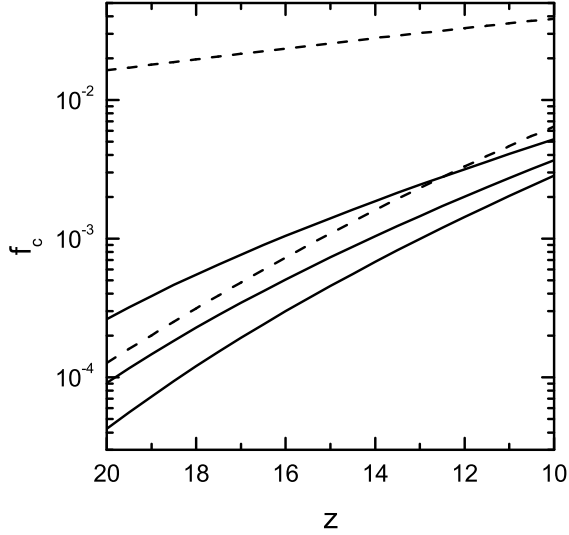}}
      \caption{
        Fraction of baryons $f_c$ in
        the universe cooled below 150 K. According
        to Gilden criterion these baryons can give rise to the formation of
        stars. Solid  curves show the fraction $f_c$ for $M_{\rm min}=
        10^3M_\odot$, $10^4M_\odot$, $10^5M_\odot$ (bottom to top)
        in eq. (\ref{massint}). Dashed curves 
        depict two cases when the lower limit in the r.h.s. of 
        (\ref{massint}) is equal to the virial mass $M_{\rm c}=M_v$ 
        with $T_{vir} = 400, 10^4$~K (top to bottom); the lower limit 
        $M_{\rm min}$ in the denominator of the r.h.s. of 
        (\ref{massint}) is taken for these two curves to be $10^4\msun$.
        }
  \label{Fig4}
\end{figure}

Characteristic star formation time inside the formed
fra\-gments is determined by the baryon density in there 
$t_{\rm sf}=t_J/\varepsilon$, where $t_J=\sqrt{3\pi/32G\rho}$ 
is the baryon Jeans time
in cold layers, $\varepsilon\ll 1$
is the star formation efficiency. Gas density $\rho_f$
in cold layers when they fragment
can be found from the condition $\rho_iv_c^2=k\rho_fT_{\rm CMB}/m_H$, which gives
\begin{equation}
n_f\simeq 0.02M_7^{2/3}(1+z)^3
~{\rm cm^{-3}},
\end{equation}
and
\begin{equation}
t_J=5\times 10^8M_7^{-1/3}(1+z)^{-3/2}~{\rm yr},
\end{equation}
For star formation efficiency $\varepsilon>0.03$ the charactersitsic time
$t_{\rm sf}$ remains shorter that the Hubble time for the halo masses
$M_h>10^7~M_\odot$. This means that star formation rate in merging halos is
determined by the longest time -- the charactersitic time between subsequent
mergings. In these conditions star formation rate is proportional to the merger
rate of the halos with the total mass above the critical value $M_{\rm c}(z)$ 
(Barkana \& Loeb 2000; Santos, Bromm \& Kamion\-kows\-ki 2002)
\begin{eqnarray}
\label{sfrate}
\dot M_\ast={1\over2}{\Omega_b\over\Omega_m}\varepsilon \epsilon_\perp
\int\limits_{M_{min}}^{M_{c}}dM_1 F(M_1) \nonumber\\
\int\limits_{M_{c}}^{M_{c}+M_1}dM_2 q_f(M_2,z)M_2 {d^2P\over dM_2 dt},
\end{eqnarray}
where
$P = P(M_1,M_2,z)$ is the probability that
a halo with mass $M_1$ merges to a halo of mass $M_2>M_1$ at redshift $z$
(Lacey \& Cole 1993);
we explicitly assume here that only fraction of baryons
$q_f(M,z)$ cooled after merging below 200 K is able to form stars.
Therefore, if
we substitute here the fraction $q_{T\leq 150}(M)$ of baryons cooled to $T<150$ K,
equation (\ref{sfrate}) will describe the contribution to the total
star formation rate from the merging halos where thermodynamics 
is governed by HD cooling, and where low mass stars may form.
It is obvious that mergings of halos of masses $M_{\rm min}\ll M_{c}$ with the halo
of critical (or overcritical) mass, involve too small baryon
mass fraction into sufficiently strong
compression where HD molecules can cool gas. Moreover, in this limit
the compressed region deviates significantly from
planar geometry, and Gilden criterion is not applicable.
Therefore, in our estimates of star formation rate we assume
for $M_{\rm min}$ two values: $M_{\rm min}=0.5~M_{c}$ and
$M_{\rm min}=0.9~M_{c}$ 
with $M_{c}$ defined as above as the mass where the fraction of 
cold baryons ($T\leq 150$ K) after merging is 
equal to 1\%; at equal 
densities of merging halos their sizes differ in the first case by $\simeq$ 
20\%, while in the second only by 4\%. Therefore, in the estimates of 
the star formation rate with a predominance of HD in thermodynamics, we 
neglect contribution from mergers of masses significantly smaller than 
the critical mass. On the other hand, it is seen from Fig. 3 that halos 
with $\sim 2M_{c}$ have already $\sim 50$\% of their mass colder than $T\leq 150$ K and unstable in 
Gilden sense. This means that the upper limit $2M_{c}$ in the second 
integral of (\ref{sfrate}) counts practically all halos whose star formation is regulated by HD cooling. 
With this proviso, the two cases: $M_{\rm min}=0.5~M_{c}$ and
$M_{\rm min}=0.9~M_{c}$,
are shown in Fig. 5 by two solid lines --
the region between them can be a reasonable estimate of the star formation
rate governed by HD cooling.

For comparison in Fig.~5 we add two lines for which $M_c$ in 
(\ref{sfrate}) is replaced by the virial mass $M_{v}$ with
$T_{vir} = 400, 10^4$~K (see, Barkana \& Loeb 2000).
It is obvious, that the number of mergers where HD cooling dominates, is only
a small (although not negligible) fraction of all mergers:
at $z$ between 10 and 16 star formation governed by HD varies from 15 to
30 \% of the one connected with 10$^4$~K halo mergers (the lower dashed line), and
0.5 to 10 \% of star formation in 400~K mergers (the upper dashed line);
in Fig. 5 for all low-mass mergers dominated by H$_2$ cooling
$q_f(M,z)=q_{f,{\rm H}_2} = 0.08$ is assumed following (Abel et al. 1998).
At earlier stages, $z=18-20$, mergers with a predominance
of HD cooling contribute less than 0.5\% compared to the 400~K mergers.
From this point of view one can think that in numerical simulations the
regions with HD cooling are apparently missed.

\begin{figure}
  \resizebox{\hsize}{!}{\includegraphics{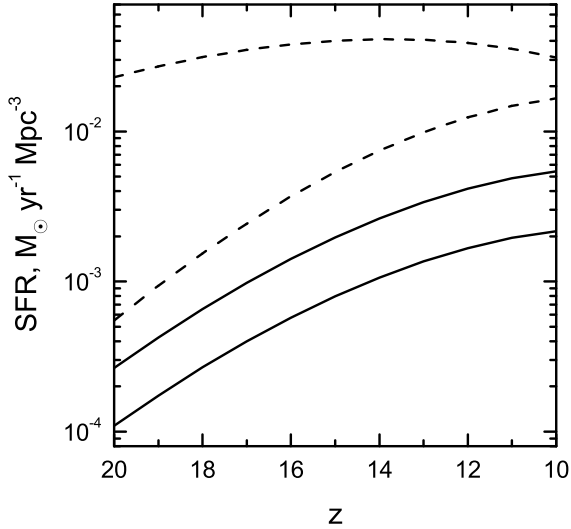}}
      \caption{
      Cosmic star formation rate:
      dashed lines correspond to the halos with
      $T_{vir} = 400, 10^4$~K (top to bottom);
      solid -- the halos with mass $M_{cr} = M_{c}$,
      $M_{\rm min} = 0.5M_{c}$ (upper curve) 
      and $M_{\rm min} = 0.9M_{c}$
      (lower curve); in all cases $\varepsilon=0.1$.
      }
  \label{Fig5}
\end{figure}

It is therefore seen that a small fraction of baryons in mergers cools
down to the lowest possible temperature $T\simeq T_{\rm CMB}$, and can give
rise to formation of the first generation stars of low masses -- lower than
the masses formed under the
conditions when H$_2$ controls thermal evolution of baryons.
This fraction increases with the total mass of merging halos, and
thus in massive galaxies the population of low-mass stars of the first generation
can be considerable. One should stress though that massive galaxies,
formed in the hierarchical scenario through mergers,
are quite expected to have been already experienced star formation episodes, with
the insterstellar gas polluted by metals. However, it remains unclear whether
the metals can become well mixed in a galaxy before it absorbs a new halo
in next merger event. Simple arguments suggest the opposite.
Indeed, the characteristic mixing time for the whole galaxy
can be estimated as $t_{mx}\sim \langle \delta_{ej}\rangle R/c_s$, where
$\langle \delta_{ej}\rangle>1$ is the mean density contrast between SNe ejecta
and diffuse interstellar gas, $R$ is the galaxy radius, $c_s$, the sound speed
in diffuse gas; note that $R/c_s\sim t_c$.
The characteristic time scale for the halo mass
growth  $t_m = [M_2 d^2 P(M_1,M_2,z)/ \\ dM_2dt]^{-1}$. 
For the halos
$M_2 \sim 10^7~M_\odot$ and $M_1\sim 0.9 M_2$ at $z =20$, $t_m$ is about
$\sim 8\times 10^{14}~$s, which is comparable to the collision time
$t_c = 3R/2v_c$, and is therefore $<t_{mx}$. Although some of the absorbed
low mass halos can have been also experienced star formation episodes before
being merged, and thus can be already metal enriched (Scannapieco, Schneider
\& Ferrara 2003),
however a non-negligible fraction of them may have pristine composition
due to a slow mixing.

The existence of low mass Pop III stars is suspected from the
observational point of view: recently discovered extremely metal-poor and
low-mass stars, as for instance a $0.8~M_\odot$ star
HE 0107-5240 (Christlieb et al. 2002, 2004)
with [Fe/H]$=-5.3$, may be the first-generation stars. The question of whether
stars with [Fe/H]$<-5$ are indeed Population III stars is still
under discussion, partly because of overabundant carbon and nitrogen:
[C/Fe]$=4$, [N/Fe]$=2.3$ in HE 0107-5240 (Bessel, Christlieb \& Gustafsson 2004).
From this point of view HE 0107-5240 can be a Population II star formed in an 
already enriched interstellar gas (Umeda \& Nomoto 2003).
However, the possibility that this star has been formed of pristine gas cannot be
excluded: Shigeyama Tsujimoto \& Yoshii (2003)
conclude that HE 0107-5240 is a Pop III
star with the surface polluted by accreted interstellar gas already enriched
with metals. In this scenario overabundant C and N in the envelope
can be produced during the core helium flash as suggested
by (Weiss et al. 2000; Schlattl et al. 2002).
The estimates presented in this paper show
that a small fraction of baryons processed by shocks in merger events may
indeed become extremely cold and form low mass stars of the first generation.

\section{Conclusion}

In this paper we have shown that

\begin{itemize}

\item

A small fraction of baryons in merging halos can cool down to very
low temperatures close to the temperature of the cosmic microwave
background. This fraction increases with the halo mass, and can reach $\simeq 0.2$
for masses $M>2\times 10^7~M_\odot$ at $z=20$ and for
$M>10^8~M_\odot$ at $z=10$; averaged over the
halo mass spectrum this fraction varies from 10$^{-3}$ at $z=20$ to 0.1 at
$z=10$.

\item

Such cold gas is unstable against gravitational fragmentation,
with the mass of the primary fragments decreasing for mergers of higher
masses $M$: $M_J\leq 1.3\times 10^4 \\ M^{-1/3}M_\odot$. Masses of protostars
formed in gas cool\-ed by HD molecules are in general lower than those
formed in conditions when H$_2$ cooling dominates.

\item

The contribution to the cosmic star formation rate of the mergers with a
predominance of HD cooling, and therefore with presumably low mass first
stellar objects, increases from less than 0.5\% at redshift $z=18-20$ to
10-30 \% at $z=10$. Extremally metal-poor low mass stars in the
Milky Way may have been formed in mergers dominated by HD cooling.

\end{itemize}

\section{Acknowledgments}

EV thanks E. Ripamonti for useful discussions. We ack\-now\-le\-dge critical remarks by the anonymous referee. This work is
supported by the Federal Agency of Education (project code RNP 2.1.1.3483),
by the RFBR (project code 06-02-16819)
and by the Rosnauka Agency grant No 02.\-438.\-11.\-7001.

\end{document}